\def\year{2023}\relax
\documentclass[letterpaper]{article} 
\usepackage{aaai23}  
\usepackage{times}  
\usepackage{helvet}  
\usepackage{courier}  
\usepackage[hyphens]{url}  
\usepackage{graphicx} 
\usepackage{natbib}  
\usepackage{caption} 
\DeclareCaptionStyle{ruled}{labelfont=normalfont,labelsep=colon,strut=off} 
\usepackage{algorithm}
\usepackage{algorithmic}

\usepackage{newfloat}
\usepackage{listings}
\floatstyle{ruled}
\newfloat{listing}{tb}{lst}{}
\floatname{listing}{Listing}
\nocopyright
\usepackage{float}

\usepackage{subfig}

\usepackage{enumitem}

\usepackage{booktabs} 

\usepackage{csquotes}
\usepackage{amsmath}
\usepackage{outlines}

\usepackage{siunitx}
\usepackage[usenames]{xcolor}

\def\capfirstletteraux#1#2\relax{\uppercase{#1}\lowercase{#2}}

\usepackage{marvosym}

\usepackage{graphicx}
\usepackage{graphbox} 
\graphicspath{{Figures/}}

\usepackage{array}
\newcolumntype{P}[1]{>{\centering\arraybackslash}m{#1}}
\newcolumntype{L}[1]{>{\arraybackslash}m{#1}}
\newcolumntype{R}{>{\raggedleft\arraybackslash}X}

\usepackage{xspace}
\newcommand\ie{i.\,e.\xspace}
\newcommand\eg{e.\,g.\xspace}

\newcommand\US{U.\,S.\xspace}

\usepackage{url}
\usepackage[normalem]{ulem}
\usepackage{balance}

\makeatletter
\renewcommand{\fps@figure}{htb}         
\renewcommand{\fps@table}{htb}         
\makeatother

\usepackage{tabularx}

\allowdisplaybreaks

\usepackage[shortcuts]{extdash}
\usepackage{soul}

\title{Community Notes vs.\ Snoping: \\How the Crowd Selects Fact-Checking Targets on Social Media}

\author{Moritz Pilarski,\textsuperscript{\rm 1} Kirill Solovev,\textsuperscript{\rm 1} Nicolas Pröllochs\textsuperscript{\rm 1}\\}
\affiliations{
	\textsuperscript{\rm 1}JLU Giessen, Germany \\
	moritz.pilarski@wi.jlug.de, kirill.solovev@wi.jlug.de, nicolas.proellochs@wi.jlug.de
}

\begin{document}

\maketitle

\begin{abstract}
	Deploying links to professional fact\-/checking websites (so\-/called ``snoping'') is a common misinformation intervention technique that can be used by social media users to refute misleading claims made by others. However, the real\-/world effect of snoping may be limited as it suffers from low visibility and distrust towards professional fact\-/checkers. As a remedy, Twitter recently launched its community\-/based fact\-/checking system ``Community Notes'' on which fact\-/checks are carried out by actual Twitter users and directly shown on the fact\-/checked tweets. Yet, an understanding of how fact\-/checking via Community Notes differs from regular snoping is largely absent. In this study, we empirically analyze differences in how contributors to Community Notes and Snopers select their targets when fact\-/checking social media posts. For this purpose, we collect and holistically analyze two {unique} datasets from Twitter: (a) \num{25912} community\-/created fact\-/checks from Twitter's Community Notes platform; and (b) \num{52505} ``snopes'' that debunk tweets via fact\-/checking replies linking to professional fact\-/checking websites. We find that Notes contributors and Snopers focus on different targets when fact\-/checking social media content. For instance, Notes contributors tend to fact\-/check posts from larger accounts with higher social influence and are relatively less likely to endorse/emphasize the accuracy of not misleading posts. Fact\-/checking targets of Notes contributors and Snopers rarely overlap; however, those overlapping exhibit a high level of agreement in the fact\-/checking assessment. Moreover, we demonstrate that Snopers fact\-/check social media posts at a higher speed. Altogether, our findings imply that different fact\-/checking approaches -- carried out on the same social media platform -- can result in vastly different social media posts getting fact\-/checked. This has important implications for future research on misinformation, which should not rely on a single fact\-/checking approach when compiling misinformation datasets. From a practical perspective, our findings imply that different fact\-/checking approaches complement each other and may help social media providers to optimize strategies to combat misinformation on their platforms.
\end{abstract}

%
%


\maketitle



\section{Introduction}
\label{sec:introduction}


Social media has shifted the quality control for content from trained journalists towards regular users \citep{Kim.2019}. The inevitable lack of oversight makes social media {platforms (\eg, Twitter, Facebook) vulnerable to misinformation \citep{Shao.2016,Pew.2016b,Kim.2019}.} If misinformation becomes viral, it can have detrimental consequences on how opinions are formed and on the offline world \citep{Allcott.2017,Moore.2023,Bakshy.2015,Oh.2013,Gallotti.2020,Geissler.2023,Jakubik.2023,Bar.2023}. In order to identify and eventually curb the spread of misinformation, third\-/party fact\-/checking organizations (\eg, \mbox{\url{snopes.com}}, \mbox{\url{politifact.com}}) regularly fact\-/check social media rumors \citep{Vosoughi.2018}. These fact\-/checking assessments are supposed to help users to identify misleading content \citep{Shao.2016}. Yet, a major challenge is that fact\-/checks from third\-/party fact\-/checking organizations suffer from low visibility as their websites are rarely visited \citep{Robertson.2020,Opgenhaffen.2022}. Users are oftentimes not aware of these fact\-/checks when consuming potentially misleading content on social media. Hence, the real\-/world effect of third\-/party fact\-/checks in curbing the spreading of misinformation on social media is limited \citep{Opgenhaffen.2022}.

\begin{figure*}
	\centering
	\captionsetup{position=top}
	\captionsetup{belowskip=0pt}
	\caption{(a) Example of a ``snoped'' post on Twitter with a reply tweet linking to a fact\-/check from a third\-/party fact\-/checking organization. (b) Example of a Community Note on Twitter.}
	\subfloat[Conversational fact\-/check (``snoping'')]{\includegraphics[width=.455\linewidth]{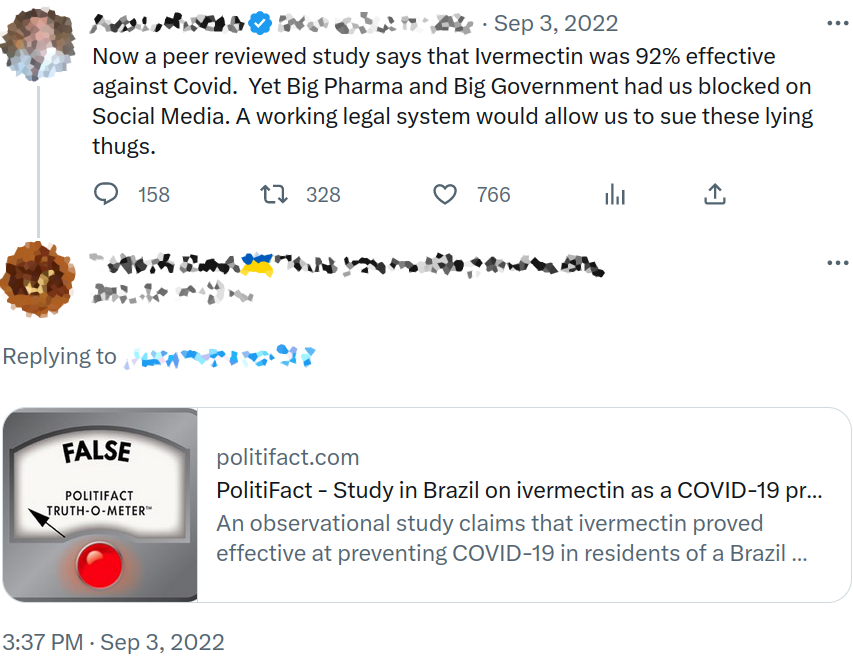}\label{fig:snope_example}}
	\hspace{1.5cm}
	\subfloat[Community Note]{\includegraphics[width=.455\linewidth]{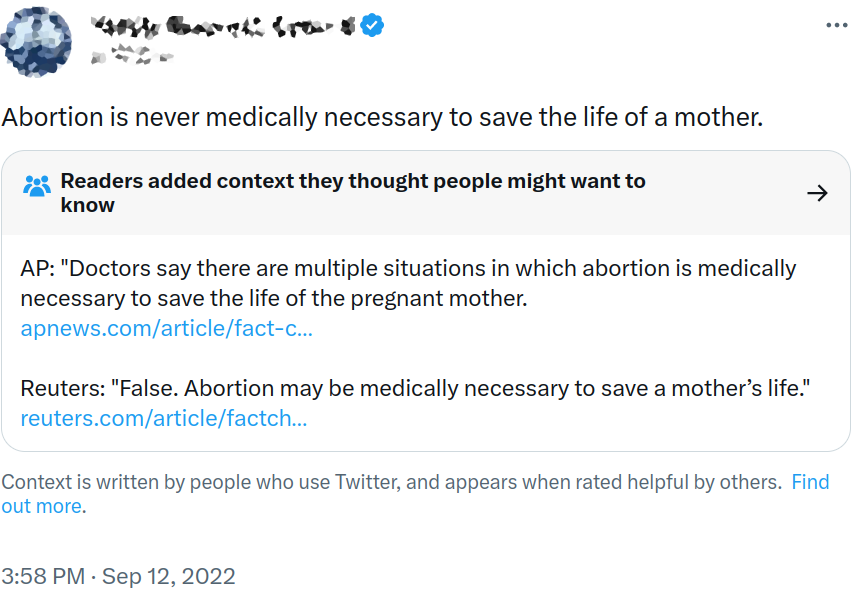}\label{fig:note_example}}
	\vspace{-.5cm}
\end{figure*}

A popular intervention to raise the visibility of third\-/party fact\-/checks on social media is conversational fact\-/checking -- also known as ``snoping'' \cite{Hannak.2014}. Here, users independently refute misleading claims in posts by replying with a link to a third\-/party fact\-/check debunking the rumor (see example in Fig.~\ref{fig:snope_example}). This approach builds on the premise that linking to a fact\-/check directly in the place where the misinformation is circulating can make the fact\-/check more visible to users who would otherwise not actively seek out for fact\-/checks \citep{Opgenhaffen.2022}. While snoping has the potential to make users more aware of third\-/party fact\-/checks, its effectiveness may still be limited for multiple reasons: (i) fact\-/checks in replies to posts may easily be overlooked and are oftentimes simply ignored by users \citep{Hannak.2014}; (ii) Snopers have been observed to focus on specific targets (\eg, members of outgroups) and snoping may be a performative rather than deliberative act (\eg, to gain social status; \citeauthor{Hannak.2014} \citeyear{Hannak.2014}). (iii) A large proportion of social media users distrust professional fact\-/checkers \citep{Pew.2019}. Hence, even when users become aware of a snoped post, the impact of the fact\-/check may be limited due to a lack of trust \citep{Brandtzaeg.2017}.

As a remedy, Twitter recently launched its community\-/based fact\-/checking system ``Community Notes,'' formerly known as ``Birdwatch'' \citep{Twitter.2021,Prollochs.2022a}. This Twitter feature allows users to identify tweets they believe are misleading or not misleading and write (textual) notes that provide {context} to the tweet. Users can add Community Notes to \emph{any} tweet they come across on Twitter. Compared to conversational fact\-/checks, Community Notes promise increased visibility as they can appear directly on the fact\-/checked tweets (see example in Fig.~\ref{fig:note_example}). Furthermore, Community Notes are carried out anonymously and may address the trust problem with professional fact\-/checkers. Recent research yielded promising results -- suggesting that Community Notes can achieve high accuracy in fact\-/checking social media posts \cite{Wojcik.2022}. However, an understanding of how fact\-/checking on a dedicated community fact\-/checking system (such as Community Notes) differs from conversational fact\-/checking (\ie, snoping) is absent. In particular, little is known regarding how (and how fast) Notes contributors and Snopers select their fact\-/checking targets and the extent to which both features complement each other. 


\textbf{Research Goal:} In this work, we empirically analyze how contributors to Community Notes and Snopers select their fact\-/checking targets on {Twitter}
. Specifically, we address the following research questions:

\begin{itemize}
\item \textbf{(RQ1)} \emph{How do the fact\-/checking targets of Notes contributors and Snopers differ in terms of author, content, and engagement characteristics?}
\item \textbf{(RQ2)} \emph{Do Community Notes reach social media users faster than conversational fact\-/checks?}
\item \textbf{(RQ3)} {\emph{How do Community Notes and Snoping complement each other?}}
\end{itemize}

\textbf{Data \& Methodology:} To address our research questions, we collected two {unique} datasets from Twitter: (a)  \num{25912} community\-/created fact\-/checks from Twitter's Community Notes platform; and (b) \num{52505} snopes that debunk tweets using fact\-/checking replies linking to professional fact\-/checking websites. We extract a wide variety of author characteristics (\eg, followers), content characteristics (\eg, topics), and engagement characteristics (\eg, virality) from the fact\-/checked tweets. This allows us to holistically analyze how Snopers and Notes contributors select their targets for fact\-/checking. Furthermore, we implement regression analysis to study differences in the fact\-/checking speed and evaluate the extent to which the fact\-/checking assessments of Snopers and Notes contributors agree.

\textbf{Contributions:} We find that Notes contributors and Snopers focus on different targets when fact\-/checking {Twitter} 
content. For instance, Notes contributors tend to fact\-/check posts from larger accounts with higher social influence and are relatively less likely to endorse/emphasize the accuracy of not misleading posts. Fact\-/checking targets of Snopers and Notes contributors rarely overlap; however, those overlapping exhibit a high level of agreement in the fact\-/checking assessment. Moreover, we demonstrate that Snopers fact\-/check {tweets} 
at a higher speed. In sum, our findings imply that different fact\-/checking approaches -- carried out on the same social media platform -- can result in vastly different social media posts getting fact\-/checked. This has important implications for future research on misinformation, which should not rely on a single fact\-/checking approach when compiling misinformation datasets. From a practical perspective, our findings imply that different fact\-/checking approaches complement each other and may help social media providers to optimize strategies to combat misinformation on their platforms.

\section{Background}

\subsection{Misinformation on Social Media}

Compared to most traditional mass media outlets, social media platforms have lower standards for content moderation. As user\-/generated content can often be disseminated among users without undergoing any significant third\-/party filtering, fact\-/checking, or editorial scrutiny \citep{Allcott.2017}, social media is much more vulnerable to the spread of misinformation {\citep{Shao.2016,Pew.2016b,Kim.2019,Lutz.2023}}. Several studies suggest that misleading information on social media tends to spread further, faster, deeper, and more widely than not misleading information \citep{Vosoughi.2018,Solovev.2022b,Prollochs.2021b,Prollochs.2023}. Misinformation is considered a serious threat to democracy and society, as it can contribute to a wide range of issues including, but not limited to, increased political polarization, threats to public safety, and erosion of trust in institutions \citep{Lazer.2018,Bar.2023}. Given these potential harms, there have been increasing calls to social media providers to take action to address the spread of misinformation on their platforms \citep{Donovan.2020}.

The most widespread approach to fact\-/checking is to have professional fact\-/checkers verify claims. While this expert fact\-/checking approach has been shown to be effective in numerous studies (a comprehensive review can be found in 
\citeauthor{Walter.2020} \citeyear{Walter.2020}), it still has several critical limitations. Due to the time\-/consuming nature of thoroughly investigating claims \citep[often many hours or even days for a single claim;][]{Guo.2022} and the limited number of fact\-/checkers available, many misleading stories never get tagged. Limited resources often force fact\-/checkers to prioritize content that is blatantly false or deliberately misleading over content that is more nuanced or complex \citep{Pennycook.2019}. As a result, they may overlook biased or misleading coverage of events, incomplete information, or the use of misleading statistics. Furthermore, many U.S.\ citizens distrust professional fact\-/checkers. According to a study conducted by the \citet{Pew.2019}, a majority of Republican partisans (\SI{70}{\percent}) and half of all U.S.\ adults believe that fact\-/checkers are biased and that their corrections cannot be trusted. Also, professional fact\-/checks oftentimes have very limited reach. Besides some collaborations with social media platform providers on specific topics, fact\-/checking organizations mainly communicate the results of their fact\-/checks through their websites. According to a study \citep{Robertson.2020}, in 2017, over half of all U.S.\ adults had never visited any fact\-/checking website. 

\subsection{Conversational Fact\-/Checking (``Snoping'')}

A popular strategy employed by social media users to combat misleading statements is to link professional fact\-/checking articles from third\-/party fact\-/checking organizations (\eg, \url{snopes.com}, \url{politifact.com}) in replies to the original message. This conversational approach to fact\-/checking -- commonly referred to as ``snoping'' \citep[\eg,][]{Hannak.2014,Friggeri.2014} -- (partially) addresses the issue of limited reach of third\-/party fact\-/checking websites by increasing the visibility of articles in the contexts of the respective fact\-/checked statements. Researchers have utilized data on conversational fact\-/checks to investigate various phenomena surrounding the spread of misinformation on social media platforms like Twitter \citep[\eg,][]{Hannak.2014,Margolin.2018,Vosoughi.2018,Mosleh.2022}, Facebook \citep[\eg,][]{Friggeri.2014}, or Reddit \citep[\eg,][]{Bond.2023}. For instance, \citet{Friggeri.2014} study the effect of snoping on the propagation of rumors on Facebook. The authors find that snopes on individual reshares of rumors increase the probability of those reshares being deleted. 

Only a few works have studied how users engage in snoping and their motifs. For example, \citet{Hannak.2014} and \citet{Margolin.2018} focus on the effect that social relations between Snopers and Snopees have on the recognition of corrections on Twitter. They find that while only a small share of all snopes is made by friends (\ie, mutually following users), those are especially likely to get the Snopee's attention. They attribute this to the circumstance that individuals feel more obligated toward their friends to respond scientifically and be more open toward facts that challenge their original positions. {Another work has studied the role of linguistic and engagement features \citep{Ma.2023}. The authors find that misinformative tweets expressing negative emotion and impoliteness are more likely to receive countering replies from users. Additionally, they observe that countered tweets tend to have a higher proportion of reply engagement compared to like, retweet, and quote tweet engagement.} Furthermore, research has analyzed the network of follower\-/relations among Snopers and Snopees. \citet{Hannak.2014} find that the network exhibits strong polarization between two large densely connected communities that roughly reflect the political camps forming along U.S.\ party lines. At the same time, most of the snope\-/relations are spanning between those communities, which suggests that snoping is commonly directed outwards as criticism of individuals that Snopers otherwise are not interested in. Hence, snoping, in many cases, should be seen as a performative rather than a deliberative act, in which Snopers are displaying their political affiliation \citep{Hannak.2014}. Since a large share of people in the U.S.\ dislikes and distrusts users from the opposing party -- a phenomenon typically subsumed under the term ``affective polarization'' \citep{Iyengar.2019} -- many snopes may go unheard.

\subsection{Community\-/Based Fact\-/Checking Systems}

As snoping essentially passes on judgments made by professional fact\-/checkers, it faces many of the same challenges and limitations as professional fact\-/checking. For example, users can not snope claims that have not yet been verified by professional fact\-/checkers. Furthermore, snoping suffers from low visibility and distrust towards fact\-/checkings organizations. A possible remedy to those problems is crowd\-/sourcing the fact\-/checking process. This would have the advantage that an abundance of users willing to participate in content moderation would grant the effort almost unlimited resources \cite{Allen.2021,Pennycook.2019}. Furthermore, trust issues with professional fact\-/checkers could be mitigated. 
Despite those promises, there are, however, reasons to also be concerned about crowd judgments. For example, unlike professional fact\-/checkers, the crowd typically lacks specific training and systematic practices for evaluating the veracity of stories \citep{Graves.2017}.

In recent years, a growing body of research \citep{Micallef.2020,Bhuiyan.2020,Pennycook.2019,Epstein.2020,Allen.2020,Allen.2021,Godel.2021,Drolsbach.2023a, Drolsbach.2023b,Prollochs.2022a} has focused on community\-/based fact\-/checking systems that leverage the ``wisdom of crowds'' \citep{Surowiecki.2005}. These systems rely on the principle that while individual users' fact\-/checks may be prone to bias or inaccuracies, high levels of accuracy can be attained through the collective judgments of politically diverse groups \citep[a summary of related literature can be found in][]{Martel.2023}. For example, \citet{Allen.2021} compare the correlation between the average ratings of differently sized crowds of laypeople and three professional fact\-/checkers to the correlation between the individual fact\-/checkers' ratings. Whereas the laypeople were merely presented the headline and lede of articles, the fact\-/checkers were thoroughly researching them. As they keep increasing the crowd size, they stop finding a significant difference between the correlations of ratings at a crowd size of about eight people \citep[similar results in][]{Bhuiyan.2020, Resnick.2021}.

Informed by those promising research findings, Twitter recently introduced its community\-/based fact\-/checking system ``Community Notes'' (formerly known as ``Birdwatch''). This new feature provides users with the ability to fact\-/check any tweet they come across by creating so\-/called Community Notes. Community Notes consist of a categorization of whether a tweet might or might not be misleading and an open text field (max.\ \num{280} characters) that allows contributors to explain their decision and include links to relevant sources. After a note is created, other users can rate its helpfulness, and if the note reaches a certain level of helpfulness, it is displayed prominently beneath the original tweet (see Fig.~\ref{fig:note_example}). Until recently, the Community Notes feature was in pilot phase and only available to registered participants in the U.S. The pilot phase started on January 25, 2021, and ended on October 6, 2022. As of December 11, 2022, registration for Community Notes is open to users worldwide, and helpful notes are visible to everyone on Twitter.

Given the recency of the platform, research on Community Notes is scant. Early works suggest that politically motivated reasoning might pose challenges in community\-/based fact\-/checking \citep{Allen.2022,Prollochs.2022a}. For instance, \citet{Allen.2022} have found that Note contributors tend to focus their fact\-/checking efforts on content posted by individuals with whom they hold opposing political views. Notwithstanding, community\-/created fact\-/checks on Community Notes have been found to be perceived as informative and helpful by the vast majority of social media users \citep{Prollochs.2022a}. {\citet{Saeed.2022} additionally highlight the important role played by Notes contributors in refuting false claims that have already been fact\-/checked by professional journalists but continue to circulate on Twitter nonetheless.
} 
Furthermore, recent research indicates that community fact\-/checked misleading posts are less viral than not misleading posts \citep{Drolsbach.2023a,Chuai.2023} and that displaying notes may reduce users' propensity to share misleading posts \citep{Wojcik.2022}. In a recent user study conducted directly on Twitter \citep{Wojcik.2022}, users were randomly assigned to view either Tweet annotations or no annotations. The results showed that those who were exposed to annotations on tweets were \SI{25}{\percent}-\SI{34}{\percent} less likely to like or retweet them compared to the control group.

\section{Data Sources}

\subsection{Dataset I: Community Notes}\label{sec:data_note}

\textbf{Fact\-/Checks:}
Community Notes is a community\-/based fact\-/checking system that allows registered users to fact\-/check statements made on Twitter. Users can fact\-/check \emph{any} tweet they come across on Twitter -- directly when browsing the platform. We obtained the data on Community Notes from the complete database dumps that are published by Twitter on a weekly basis.\footnote{\url{https://twitter.com/i/birdwatch/download-data}} From this dataset, we used the notes' publication dates, veracity judgments (\ie, whether the tweet is categorized as misleading or not misleading), as well as the free\-/text explanations (max.\ \num{280} characters) that are used by contributors to explain their judgments. In our study, we consider all fact\-/checks that were created during Community Note's pilot phase in the \US, which started on January 26, 2021 and ended on October 5, 2022.

\textbf{Fact\-/Checked Tweets:}
We used Twitter's tweet lookup API endpoint to collect all fact\-/checked tweets, \ie, tweets that have received a Community Note. Furthermore, we collect various information about the authors of the fact\-/checked tweets (\eg, number of followers, verified status). We excluded all tweets that were not classified as written in English by Twitter's language detection algorithm as well as all tweets by the user \texttt{@CommunityNotes} since those were officially recommended for testing purposes. Notably, multiple contributors can write Community Notes for the same tweet. Therefore, the data sometimes includes multiple fact\-/checks for the same post. In our data, \SI{22.0}{\percent} of the fact\-/checked tweets received more than one Community Note. {Our final dataset encompasses a total of \num{25912} Community Notes, contributed by \num{4288} unique (pseudonymous) contributors, covering a total of \num{18805} distinct tweets (Dataset I).}
{All of our data was collected in late February 2023. Any content that was deleted before that time is not included in our analysis.}

We performed basic text preprocessing on the fact\-/checked tweets by removing user\-/mentions (\texttt{@screen\-/name}) from the beginnings of the tweets' texts\footnote{User mentions at the beginning of a tweet typically refer to the structures of the reply\-/trees in which the tweets are embedded.}, removing URLs, and parsing HTML\-/characters (\eg, \texttt{\&amp;} $\rightarrow$ \&). 

\subsection{Dataset II: Conversational Fact\-/Checks (``Snopes'')}

\textbf{Fact\-/Checks:} Our approach to collecting conversational fact\-/checks, \ie, snopes, was guided by best practices from earlier research \citep{Vosoughi.2018}. We focused on three reputable fact\-/checking websites that thoroughly investigate social media rumors, namely, \url{snopes.com}, \url{politifact.com}, and \url{truthorfiction.com}. We scraped all fact\-/checks and their corresponding veracity judgments published on any of these websites (a total of \num{44086} articles). The fact\-/checking organizations have different ways of labeling the veracity of a story. For example, \url{politifact.com} articles are given a ``Pants on Fire'' rating for misleading stories, whereas snopes.com assigns a ``false'' label. Analogous to earlier work \citep{Vosoughi.2018,Solovev.2022b}, we normalized the veracity labels across the different sites by mapping them to a score of 1 to 5. All stories with a score of 1 or 2 were categorized as ``misleading,'' whereas stories with a score of 4 or 5 were categorized as ``not misleading'' (\eg, ``Pants on Fire!'' $\rightarrow$ \emph{misleading}).\footnote{For the sake of simplicity and comparability, we omitted stories with a score of 3, \ie, stories with a ``mixed'' veracity (\SI{10.5}{\percent} of all conversational fact\-/checks). Including those stories yields qualitatively identical results in our later analysis.}

\textbf{Fact\-/Checked Tweets:} We used Twitter's full\-/archive search API endpoint to collect all reply tweets featuring a link to any of the previously scraped fact\-/checking articles. To ensure comparability with Dataset I, we considered only replies that were posted between January 25, 2021 and October 6, 2022 (\ie, during Community Note's pilot phase). {As mentioned earlier, all the data we collected is from late February 2023. Any content deleted prior to that date is not accounted for in our analysis.}	
Of those tweets, we excluded all that were not classified as written in English by Twitter's language detection algorithm. To ensure that replies featuring links to fact\-/checking articles are actual fact\-/checks of statements made in their respective parent tweets, we compared the semantic contents of the fact\-/check articles' assessed claims with the texts of the tweets to which they were given as replies. Given that \SI{18}{\percent} of the fact\-/checked tweets have images attached, we first employed optical character recognition to extract the textual content from those images\footnote{We preprocessed the images with ImageMagick (\citeyear{ImageMagick}), performed optical character recognition with the Tesseract engine \citep{Smith.2007}, and performed several postprocessing steps based on DBSCAN clustering to identify coherent lines of text.}. After applying the same preprocessing steps as before, we generated document embeddings for all fact\-/checked tweets' texts (including the ones retrieved from the images) and all fact\-/checking articles' assessed claims using the pre\-/trained TwHIN-BERT language model \citep{Zhang.2022}. Finally, we calculated cosine\-/similarities between the embedding\-/vectors of all observed pairs of text and discarded those with a similarity\-/value below \num{0.75}. 
This resulted in a final dataset {comprising \num{52505} conversational fact\-/checks contributed by \num{34188} unique authors, covering a total of \num{45368} unique tweets (Dataset II).} 

\textbf{User Study:} We evaluated the performance of our method for excluding unrelated pairs of snopes and tweets with a user study. To this end, we employed two trained research assistants (hourly wage: $\approx$\$14) that were tasked with rating the semantic similarity of tweets with the corresponding fact\-/checked claim. For this, participants had to answer the question ``How related is this tweet to the fact\-/checked claim?'' on a 5\-/point Likert scale ranging from ``Completely Unrelated'' to ``Completely Related.'' We observed a relatively high Kendall's coefficient of concordance of $W$ = \num{0.738} ($p$ $<$ \num{0.001}), and \SI{81.4}{\percent} of the pairs classified as related by our model were adjudged to be at least ``somewhat related'' by the human raters. This implies that our method identifies snoped posts on Twitter with high accuracy. 

\section{Empirical Analysis}

\subsection{Target Selection (RQ1)}

To answer \textbf{RQ1}, we analyze how the fact\-/checking targets of Notes contributors and Snopers differ in terms of their account, content, and engagement characteristics.

\begin{figure}[!h]
	\centering
	\includegraphics{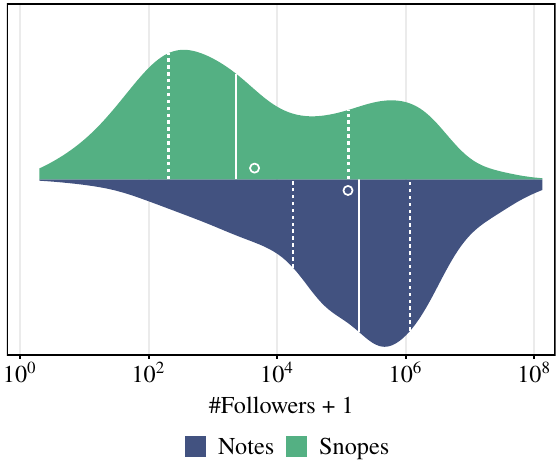}
	\caption{{Split violin plot comparing the distributions of follower counts among authors of fact\-/checked tweets. Shown are kernel density estimates (colored areas), mean values (white circles), and quartile values (white lines).}}
	\label{fig:annotation_user_feature_count_violin_plot}
\end{figure}

\textbf{Account Characteristics:} 
{Fig.~\ref{fig:annotation_user_feature_count_violin_plot} plots the kernel density estimates as well as mean and quartile values for the distributions of the fact\-/checked users' numbers of followers. We find that Notes contributors tend to annotate tweets authored by users with much higher popularity and reach. The mean number of followers for notes is almost five times higher than it is for snopes (mean\textsubscript{notes} = \num{3385810}; mean\textsubscript{snopes} = \num{679684}; [KS-test: $D$ = \num{0.409}; $p$ $<$ \num{0.001}]).}

\begin{figure}[!b]
	\includegraphics{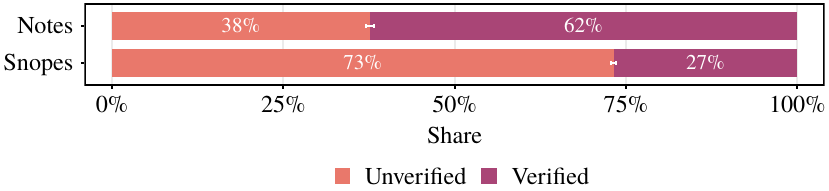}
	\caption{{Proportions of fact\-/checking targets that are verified users and their \SI{95}{\percent} confidence intervals (white error bars).}}
	\label{fig:annotation_user_is_verified_plot}
\end{figure}
{Figure \ref{fig:annotation_user_is_verified_plot} presents the distribution of fact\-/checks across tweets posted by users whose account authenticity has been verified by Twitter, potentially indicating greater perceived credibility in their statements. Notably, a significantly larger proportion of Notes contributors (\SI{62.4}{\percent}) compared to Snopers (\SI{26.8}{\percent}) focus their fact\-/checking efforts on tweets by verified users ($\chi^2$-test: $X^2$ = \num{9269}; $p$ $<$ \num{0.001}).}

{We further analyzed additional account characteristics such as the users' followee counts and account ages. Here we found comparably smaller differences. On average, Notes contributors are slightly more likely to focus on users with higher followee counts and older user accounts (see Supplementary Materials for details).}

\begin{figure}[h]
	\centering
	\includegraphics{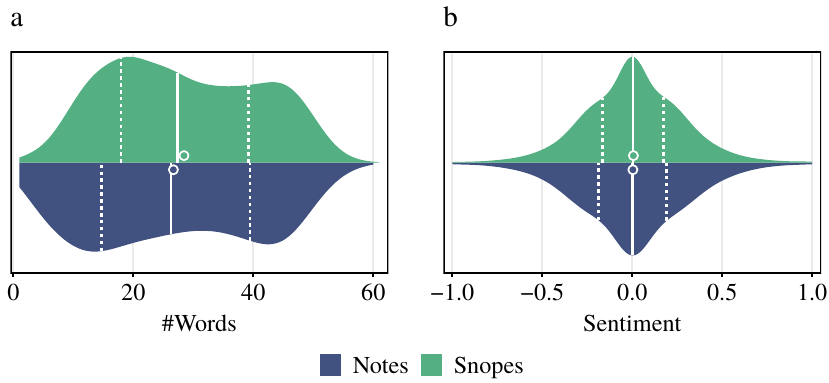}
	\caption{{Split violin plot comparing the distributions of the (a) lengths and (b) sentiment scores of the fact\-/checked tweets. Shown are kernel density estimates (colored areas), mean values (white circles), and quartile values (white lines).}}
	\label{fig:annotation_sentiment_word_count_violin_plot}
\end{figure}

\textbf{Content Characteristics:} We determined the number of word tokens (\#Words) and calculated sentiment scores based on the NRC Word\-/Emotion Association Lexicon \citep[EmoLex;][]{Mohammad.2010} for all fact\-/checked tweets. {The number of word tokens (\ie, the tweet length) potentially indicates the extent of detail within the fact\-/checked claims, while the expressed sentiment might affect the readers' emotional reactions towards the statements.} For our sentiment analysis, we use the default implementation of the \texttt{sentimentr} R package (with the built\-/in NRC lexicon) that also accounts for negations and valence shifters (see \citeauthor{Rinker.2019} \citeyear{Rinker.2019} for details), analogous to previous research \citep[\eg,][]{Robertson.2023,Prollochs.2021a}. Fig.~\ref{fig:annotation_sentiment_word_count_violin_plot} visualizes the corresponding distributions. There is a slightly higher share of notes than snopes on relatively short tweets (mean\textsubscript{notes} = \num{26.7}; mean\textsubscript{snopes} = \num{28.4} [KS-test: $D$ = \num{0.105}; $p$ $<$ \num{0.001}]). However, there is no significant difference in the mean sentiment scores (mean\textsubscript{notes} = \num{0.004}; mean\textsubscript{snopes} = \num{0.006}; [$t$-test: $t$ = \num{-1.20}, $p$ = \num{0.228}]). Overall, the observed differences regarding the length and sentiment of the fact\-/checked tweet are rather small.\footnote{We additionally analyze discrete emotions (\eg, anger, fear) in the Supplementary Materials. Again, the observed differences between Notes contributors and Snopers are small.}

{Next, we conducted topic modeling to explore potential differences in the topics that Notes contributors and Snopers focus on. 
Our rationale is that different topics may imply distinct groups of authors and target audiences, and, thus, may draw different types of fact\-/checkers.
To this end, we employed supervised machine learning to categorize the fact\-/checked tweets from our dataset into eight predefined topics: \emph{Business; Disasters; Entertainment; Health; Politics; Science; War; Other}.}
These topics have been identified based on a manual assessment of the fact\-/checked tweets in our dataset and the selection of topics in previous works \citep[\eg,][]{Vosoughi.2018}. To create training data, we employed a trained research assistant to assign topic labels (multiple selection possible) to a random subset of \num{7500} tweets. We then used the created labeled data to train a deep neural network classifier that predicts whether a tweet belongs to each topic. The input data for the training machine learning classifier was a vector representation of the labeled tweets and the topic labels. To create vector representations of tweets, we used the pre\-/trained TwHIN\-/BERT language model \citep{Zhang.2022}. In our deep neural network classifier, we treated the task of predicting topic labels for (vector representations of) tweets as a multi\-/label problem considering that one tweet may belong to multiple topics. All hyperparameters were tuned using 10-fold cross\-/validation. Our classifier achieved a relatively high micro\-/averaged $F_1$ score of \num{0.75} and an accuracy of \num{0.93} on out\-/of\-/sample tweets.

\begin{figure}[!h]
	\includegraphics{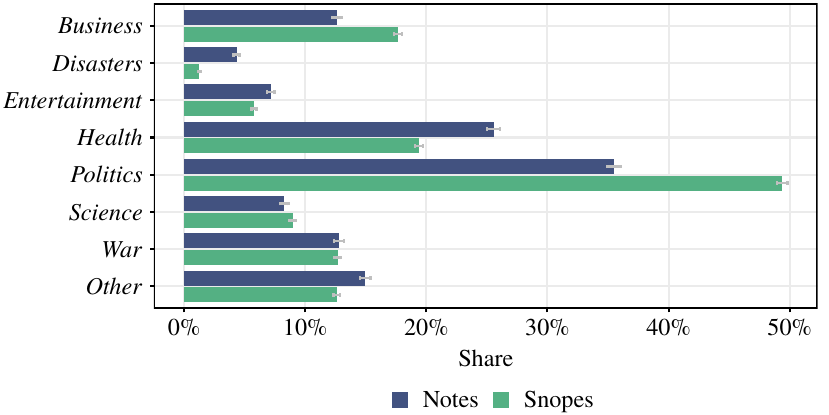}
	\caption{{Proportions of fact\-/checks on tweets with different topics (colored bars) and their \SI{95}{\percent} confidence intervals (gray error bars).}}
	\label{fig:annotation_topic_share_plot}
\end{figure}

The shares of fact\-/checks on tweets per topic are displayed in Fig.~\ref{fig:annotation_topic_share_plot}. Note that since the fact\-/checked tweets can have multiple topic labels, those shares do not sum up to \SI{100}{\percent}. There are significant differences in the distributions of the fact\-/checked tweets' topics between Community Notes and snopes ($\chi^2$-test: $X^2$ = \num{2164}; $p$ $<$ \num{0.001}). In particular, Community Notes are relatively more prevalent on tweets about \emph{Disasters}, \emph{Entertainment}, \emph{Health}, and \emph{Other} topics. In contrast, snopes are relatively more common on tweets about \emph{Business}, \emph{Politics}, and \emph{Science}.

\begin{figure}[!b]
	\includegraphics{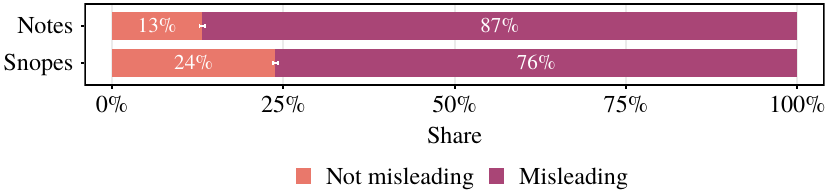}
	\caption{{Proportions of fact\-/checks' verdicts (colored bars) and their \SI{95}{\percent} confidence intervals (white error bars)}.}
	\label{fig:annotation_verdict_bar_plot}
\end{figure}

Furthermore, we examine differences in veracity judgments between Snopers and Notes contributors (see Fig.~\ref{fig:annotation_verdict_bar_plot}). Our analysis reveals that a majority of Snopers and Notes contributors adjudge the claims made in their targeted tweets as misleading. Specifically, for notes, the proportion of misleading verdicts (\SI{86.8}{\percent}) is \num{6.6} times higher than that of not misleading verdicts (\SI{13.2}{\percent}). Snopers classify \num{3.2} times more of the fact\-/checked tweets as misleading (\SI{76.2}{\percent}) than as not misleading (\SI{23.8}{\percent}). Overall, snopes exhibit a relatively higher share of not misleading verdicts compared to Community Notes [$\chi^2$-test: $X^2$ = \num{1215}; $p$ $<$ \num{0.001}].

\begin{figure}[!h]
	\includegraphics{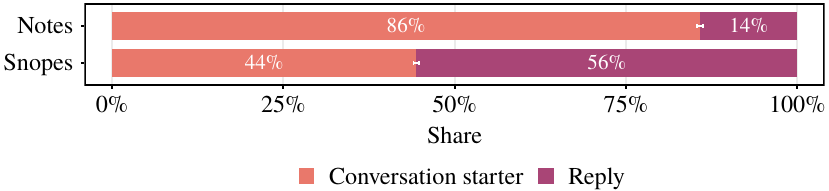}
	\caption{{Proportions of fact\-/checks on tweets that are either conversation\-/starters or replies (colored bars) and their \SI{95}{\percent} confidence intervals (white error bars)}.}
	\label{fig:annotation_is_convers_start_bar_plot}
\end{figure}

{
We also examine whether fact\-/checkers show a preference for fact\-/checking conversation starting tweets or reply tweets (see Fig.~\ref{fig:annotation_is_convers_start_bar_plot}). Understanding this difference is important as conversation starting tweets usually have a higher visibility than reply tweets \citep{Hannak.2014}. Our analysis reveals that the proportion of Notes addressing conversation starting tweets (\SI{86}{\percent}) is nearly twice as high as the proportion for Snopes (\SI{44}{\percent}; $\chi^2$-test: $X^2$ = \num{12223}; $p$ $<$ \num{0.001}).
}

\begin{figure}[!b]
	\includegraphics{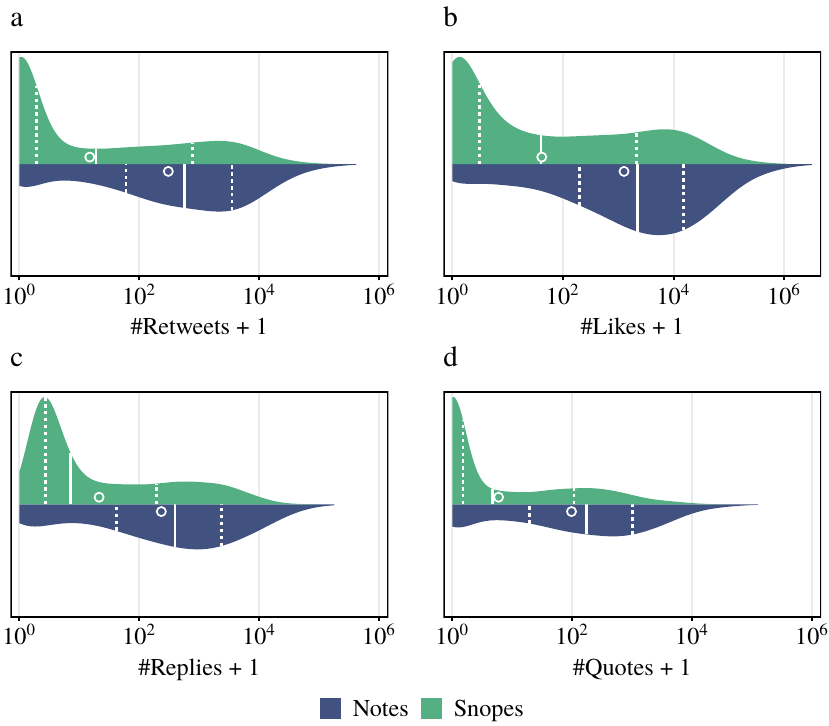}
	\caption{{Split violin plot comparing the distributions of the fact\-/checked tweets' different engagement metrics, namely, (a) the number of retweets, (b) the number of likes, (c) the number of replies, (d) the number of quotes. Shown are kernel density estimates (colored areas), mean values (white circles), and quartile values (white lines).}}
	\label{fig:annotation_tweet_feature_count_violin_plot}
\end{figure}
	
\textbf{Engagement Characteristics:} Fig.~\ref{fig:annotation_tweet_feature_count_violin_plot} depicts the distributions of the fact\-/checked tweets' engagement metrics for Community Notes and snopes. We observe much higher values for Community Notes across all dimensions. Notes contributors, on average, fact\-/check tweets with almost five times more likes (mean\textsubscript{notes} = \num{28519}; mean\textsubscript{snopes} = \num{6089}; [KS-test: $D$ = \num{0.415}; $p$ $<$ \num{0.001}]), nearly four times more retweets (mean\textsubscript{notes} = \num{4816}; mean\textsubscript{snopes} = \num{1265}; [KS-test: $D$ = \num{0.425}; $p$ $<$ \num{0.001}]), roughly five times more replies (mean\textsubscript{notes} = \num{3157}; mean\textsubscript{snopes} = \num{638}; [KS-test: $D$ = \num{0.397}; $p$ $<$ \num{0.001}]), and close to 8 times more quotes than Snopers (mean\textsubscript{notes} = \num{1504}; mean\textsubscript{snopes} = \num{200}; [KS-test: $D$ = \num{0.441}; $p$ $<$ \num{0.001}]). These results suggest that Notes contributors are more likely to fact\-/check highly ``viral'' posts, whereas Snopers tend to focus on more ``regular'' posts.

\subsection{Fact\-/Checking Speed (RQ2)}

To answer \textbf{RQ2}, we analyze the lengths of the timespans between the posting dates of the original tweets and the fact\-/checks. For this purpose, we first compare summary statistics. Subsequently, we implement an explanatory regression model to analyze which tweet features are linked to a higher fact\-/checking speed.

\begin{figure}[!h]
	\centering
	\includegraphics{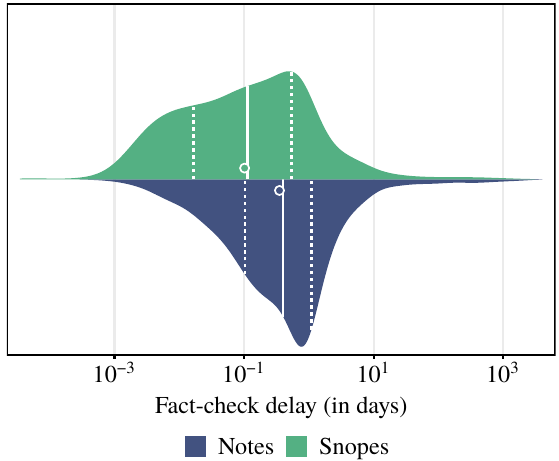}
	\caption{{Split violin plot comparing the distributions of the fact\-/checking delays, \ie, the lengths of the timespans between the posting dates of the original tweets and the fact\-/checks (in days). Shown are kernel density estimates (colored areas), mean values (white circles), and quartile values (white lines).}}
	\label{fig:annotation_time_delay_violin_plot}
\end{figure}

\textbf{Summary Statistics: } Fig.~\ref{fig:annotation_time_delay_violin_plot} shows the distributions of fact\-/check delays (time in days). The lengths of the timespans between the publication dates of fact\-/checks and their respective parent tweets tend to be longer for Community Notes than for snopes. It takes Notes contributors, on average, more than twice as long as Snopers to publish their fact\-/checks (mean\textsubscript{notes} = \SI{10.8}{days}; mean\textsubscript{snopes} = \SI{4.9}{days}; [KS-test: $D$ = \num{0.24}; $p$ $<$ \num{0.001}]).

\textbf{Regression Analysis:}
To further examine the differences in fact\-/checking delays, we perform an explanatory regression analysis. The dependent variable is the fact\-/checking delay (in days), \ie, the timespans between the posting dates of the original tweets and the fact\-/checks. The explanatory variables comprise the author, content, and engagement\footnote{The engagement characteristics (\eg, \#Retweets, \#Likes) are highly correlated. To circumvent possible multicollinearity issues, we restricted our model to \#Retweets and \#Replies.} characteristics of the fact\-/checked tweet that were presented in the previous analyses. 
We also include monthly fixed effects to control for differences in the fact\-/checking date. In our model, the fact\-/checking delays are first log\-/transformed and then modeled via a normal distribution. This modeling approach is consistent with previous research assuming a log\-/normal distribution of response times \citep[\eg,][]{Prollochs.2021b,Prollochs.2021a} and allows us to estimate the model using ordinary least squares (OLS). 
We $z$-standardized all continuous explanatory variables in order to facilitate interpretability.

{As detailed in the previous section, significant differences exist across nearly all the examined attributes' distributions between Community Notes and Snopes. In order to reduce the possibility of confounding biases in regression outcomes, we implemented propensity score matching. The propensity scores were calculated using logistic regression, followed by nearest neighbor matching with calipers set at \num{0.1} standard deviations of the propensity scores' distribution. This process culminated in a dataset encompassing \num{19,545} observations from each group. The outcome was a substantial reduction in standardized mean differences across all variables to levels below \num{0.05}, with an average relative reduction of those differences by \SI{67.4}{\percent}. In the following, we present the regression outcomes for the propensity\-/matched dataset. The results of the regression conducted on the unmatched (\ie, complete) dataset can be found in the Supplementary Materials (the results are qualitatively identical).}

\begin{figure}[tb]
	\includegraphics{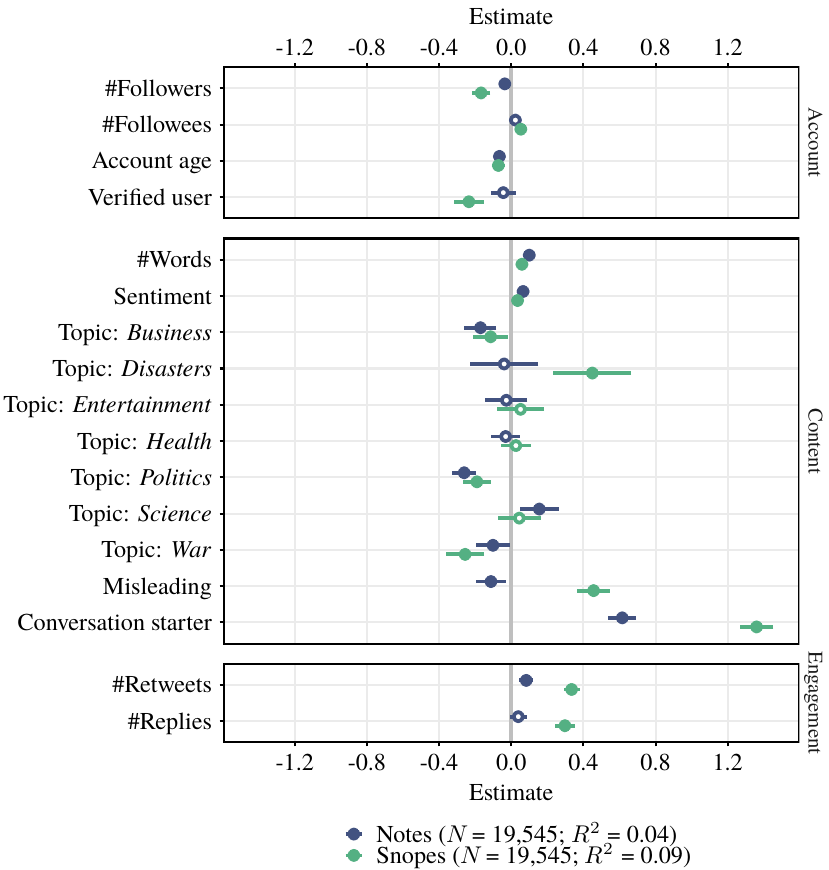}
	\caption{Coefficient estimates (circles) and their \SI{95}{\percent} confidence intervals (bars) {based on the propensity\-/matched datasets}. The dependent variable is the fact\-/checking delay, \ie, the lengths of the timespans between the posting dates of the original tweets and the fact\-/checks. Intercepts and monthly fixed effects are included. Coefficient estimates that are statistically significant ($p$ $<$ \num{0.05}) are shown with filled circles.}
	\label{fig:annotation_time_delay_model_prop_match_coef_plot}
\end{figure}

{
\textbf{Coefficient Estimates:}
Fig.~\ref{fig:annotation_time_delay_model_prop_match_coef_plot} illustrates the regression coefficients and their corresponding \SI{95}{\percent} confidence intervals. We commence by presenting the findings for the Community Notes dataset (depicted as the blue model in Figure 10). The results indicate that tweets authored by individuals with higher social influence undergo fact\-/checking at an accelerated pace. A one standard deviation increase in the number of followers corresponds to an $e^{\text{\num{-0.036}}}\approx$ \SI{3.50}{\percent} reduction in the time taken for fact\-/checking (coef.\ = \num{-0.036}, $p$ = \num{0.042}). Additionally, we observe that tweets originating from older accounts receive faster fact\-/checking. A one standard deviation increase in account age correlates with a \SI{6.34}{\percent} decrease in fact\-/checking delays (coef.\ = \num{-0.066}, $p$ $<$ \num{0.001}).
When considering tweet characteristics, we find that longer tweets (coef.\ = \num{0.100}, $p$ $<$ \num{0.001}) and those with a positive sentiment (coef.\ = \num{0.066}, $p$ $<$ \num{0.001}) tend to undergo slower fact\-/checking. Conversation starters experience a substantial \SI{85.03}{\percent} increase in fact\-/checking time (coef.\ = \num{0.615}, $p$ $<$ \num{0.001}), whereas misleading tweets exhibit a \SI{10.63}{\percent} decrease in fact\-/checking time (coef.\ = \num{-0.112}, $p$ = \num{0.007}). In terms of topics, tweets discussing \emph{Science} face \SI{16.84}{\percent} slower fact\-/checking times (coef.\ = \num{0.156}, $p$ = \num{0.005}). Conversely, tweets related to \emph{War} and \emph{Business} experience \SI{9.63}{\percent} (coef.\ = \num{-0.101}, $p$ = \num{0.038}) and \SI{15.68}{\percent} (coef.\ = \num{-0.171}, $p$ $<$ \num{0.001}) faster fact\-/checking times, respectively. Tweets involving \emph{Politics} are subjected to the fastest fact\-/checking, displaying an estimated \SI{22.96}{\percent} reduction in the time before fact\-/checking (coef.\ = \num{-0.261}, $p$ $<$ \num{0.001}).
Finally, we examine several engagement metrics, as indicated by the number of retweets and replies. A one standard deviation increase in the number of retweets leads to an \SI{8.66}{\percent} increase in fact\-/checking time (coef.\ = \num{0.083}, $p$ $<$ \num{0.001}), while the coefficient associated with the reply count does not achieve statistical significance within common thresholds.
}

{
Next, we compare the estimates with the regression results for Snopers (see the green model in Figure 10). We again observe that tweets from individuals with higher social influence tend to undergo faster fact\-/checking. In particular, a one standard deviation increase in the number of followers corresponds to approximately a \SI{15.39}{\percent} reduction in fact\-/checking time (coef.\ = \num{-0.167}, $p$ $<$ \num{0.001}). Tweets from verified users are estimated to undergo \SI{20.92}{\percent} faster fact\-/checking (coef.\ = \num{-0.235}, $p$ $<$ \num{0.001}). A one standard deviation increase in the number of followees has a slight positive effect and is associated with a \SI{5.46}{\percent} increase in fact\-/checking time (coef.\ = \num{0.053}, $p$ = \num{0.001}). 
Similar to Notes contributors, we find that longer tweets (coef.\ = \num{0.059}, $p$ $<$ \num{0.001}) and tweets with positive sentiment (coef.\ = \num{0.035}, $p$ = \num{0.030}) tend to undergo fact\-/checking at a slower pace. Conversation starting tweets exhibit a significantly higher fact\-/checking delay, with a remarkable \SI{289.61}{\percent} increase in time/delays compared to replies (coef.\ = \num{1.360}, $p$ $<$ \num{0.001}). Different from the Notes contributors model, misleading tweets tend to receive \SI{57.87}{\percent} slower fact\-/checking by Snopers (coef.\ = \num{0.475}, $p$ $<$ \num{0.001}). Examining topic effects reveals \SI{56.76}{\percent} slower (coef.\ = \num{0.450}, $p$ $<$ \num{0.001}) fact\-/checks for tweets related to \emph{Disasters}. Similar to the model for Note contributors, \emph{Business}\-/related tweets correspond to a \SI{10.77}{\percent} faster fact\-/checking (coef.\ = \num{-0.114}, $p$ = \num{0.020}). The shortest delay between a tweet and its corresponding fact\-/check for Snopers is associated with \emph{War} and \emph{Politics}, with tweets on these topics corresponding to a \SI{22.56}{\percent} (coef.\ = \num{-0.256}, $p$ $<$ \num{0.001}) and \SI{17.34}{\percent} (coef.\ = \num{-0.190}, $p$ $<$ \num{0.001}) reduction in time before fact\-/checking, respectively. 
The engagement metrics show similar effects as for the Notes contributors. A one standard deviation increase in the number of retweets is associated with a \SI{39.76}{\percent} longer delay until fact\-/checking (coef.\ = \num{0.335}, $p$ $<$ \num{0.001}), and a one standard deviation increase in the number of replies corresponds to a \SI{34.66}{\percent} longer fact\-/checking delay (coef.\ = \num{0.298}, $p$ $<$ \num{0.001}).
}

In summary, we observe a clear similarity in the way both groups tend to fact\-/check high\-/status individuals on Twitter relatively faster, and stark differences across different topics. {While political content is fact\-/checked quickly by both groups, Snopers exhibit a relatively longer delay in verifying the accuracy of tweets related to \emph{Disasters}. In contrast, Notes contributors take more time assessing the veracity of tweets concerning \emph{Science}.} 
Interestingly, we observe different signs for the coefficients of the veracity label in the two models. Specifically, we find that tweets considered as misleading are fact\-/checked slightly faster by Notes contributors, while they are fact\-/checked slower by Snopers in comparison to tweets considered as not misleading. Out of all explanatory variables in our models, the conversation starting status is associated with the highest difference in fact\-/checking delays for both, notes and snopes. A plausible explanation for this finding is that old replies have lower visibility than old conversation starting tweets and, thus, are less likely to get fact\-/checked at a later date. 

{
\textbf{Robustness Checks: } We conducted a wide variety of checks to validate the robustness of our analysis.
First, we carried out standard diagnostic tests to validate the fulfillment of key OLS assumptions. This encompassed a range of checks, including confirming that all variance inflation factors were well below the critical threshold of \num{4} and verifying the normality of the residuals. Second, there is a possibility of a bidirectional relationship where engagement not only determines the delay in fact\-/checking but also vice versa. To alleviate such endogeneity concerns, we repeated our analysis and excluded the engagement metrics from the regressions models. The results were qualitatively identical with no significant alterations in the magnitudes, signs, or significance values of the other coefficients.
}

\subsection{Overlap and Agreement (RQ3)}

{Next, we explore the extent to which the two fact\-/checking approaches complement each other. For this purpose, we examine the overlap and the within\-//between\-/group agreement in the fact\-/checking assessments of Notes contributors and Snopers (\textbf{RQ3}).}

\textbf{Overlap:} To analyze the overlap between contributors to Community Notes and Snopers, we map the tweet IDs of the fact\-/checked tweets in Dataset I to those in Dataset II. We find that \SI{28.7}{\percent} (\num{18224}) of all fact\-/checked tweets are exclusively fact\-/checked by Notes contributors, \SI{70.4}{\percent} (\num{44787}) are exclusively fact\-/checked by Snopers, and merely \SI{0.9}{\percent} (\num{581}) are fact\-/checked by both groups. Overall, this implies that the fact\-/checking targets of Snopers and Notes contributors rarely overlap.

\begin{figure}[!h]
	\centering
	\includegraphics{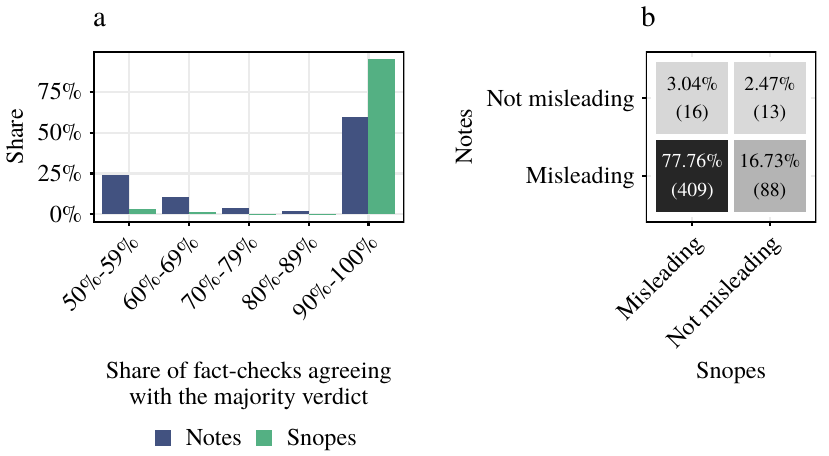}
	\caption{(a) Shares of fact\-/checks per group that agree with the majority verdict for tweets with multiple fact\-/checks (\ie, within\-/group agreement). (b) Agreement of majority verdicts between Notes contributors and Snopers for tweets have have been fact\-/checked by both groups (\ie, between\-/group agreement).}
	\label{fig:tweet_agreement_class_plot}
\end{figure}

\textbf{Within\-/Group Agreement:} Community fact\-/checkers sometimes create multiple fact\-/checks for the same post. This allows us to study the within\-/group agreement of the fact\-/checking verdicts (\ie, whether the tweet is categorized as misleading or not misleading). Among the tweets with any Community Notes, \SI{22.0}{\percent} (\num{4131}) have multiple notes associated with them. In contrast, among the tweets with snopes, only \SI{6.3}{\percent} (\num{2854}) have multiple snopes associated with them. This suggests that Notes contributors tend to concentrate their efforts on a narrower set of targets, while Snopers exhibit a broader coverage.  Fig.~\ref{fig:tweet_agreement_class_plot}a shows the distributions of the shares of fact\-/checks agreeing with the respective majority verdicts. Among the tweets with Community Notes, the average share of agreement with the majority verdict is \SI{83.1}{\percent}. On the other hand, tweets with snopes show a higher average agreement of \SI{97.9}{\percent} (KS-test: $D$ = \num{0.355}; $p$ $<$ \num{0.001}). This discrepancy may be attributed to the fact that snopes rely on verdicts from professional fact\-/checking organizations, which typically exhibit a very high level of agreement \citep{Vosoughi.2018}.

\textbf{Between\-/Group Agreement:} Fig.~\ref{fig:tweet_agreement_class_plot}b shows the agreement between the majority verdicts of Notes contributors and Snopers for all tweets that have been fact\-/checked by both groups. The overall agreement share between Notes contributors and Snopers is high at \SI{80.2}{\percent}. Notably, we observe a much higher between\-/group agreement for tweets considered \emph{misleading} compared to tweets considered \emph{not misleading}. However, the findings for the latter should be interpreted with caution due to the limited number of tweets (\num{117}) with overlapping fact\-/checks and a not misleading majority verdict by either Notes contributors or Snopers.

\section{Discussion}

\textbf{Relevance:} There are widespread concerns that misinformation on social media is damaging societies and democratic institutions \cite{Lazer.2018}. Hence, policy initiatives around the world urge social media platforms to limit its spread. A crucial prerequisite to curb the spread of misinformation on social media is its accurate identification \cite{Pennycook.2019}. Community\-/based fact\-/checking has the potential to partially overcome the drawbacks of alternative approaches to fact\-/checking, \eg, in terms of speed, volume, and trust \cite{Allen.2020}. While earlier studies suggest that crowds might be {able} to accurately assess the veracity of social media content \citep{Bhuiyan.2020,Epstein.2020,Pennycook.2019}, an understanding of how community fact\-/checkers select their targets for fact\-/checking is still largely absent. Here, we contribute to research into misinformation and fact\-/checking by characterizing how contributors to Community Notes and Snopers select their targets when fact\-/checking tweets on the social media platform Twitter.

\textbf{Summary of Findings:} Our key findings are as follows: (i) The targets of Notes contributors and Snopers significantly differ in terms of their author, content, and engagement characteristics. For instance, Notes contributors tend to fact\-/check posts from larger accounts with higher social influence and are relatively less likely to endorse/emphasize the accuracy of not misleading posts (\emph{RQ1}). (ii) Compared to Notes contributors, Snopers fact\-/check {tweets} at a higher speed (\emph{RQ2}). (iii) The fact\-/checking targets of Notes contributors and Snopers rarely overlap; however, those overlapping exhibit a high level of agreement in the fact\-/checking assessment (\emph{RQ3}).

\textbf{Implications:} Our analysis implies that Notes contributors and Snopers focus on different targets when fact\-/checking {Twitter
} content. A possible reason is that these user groups have different motivations and goals when fact\-/checking {tweets
}. In previous research, Snopers have already been observed to frequently focus on specific targets such as, for example, outgroup members (\eg, to gain social status). As such, their motivation to fact\-/check social media posts may be -- at least partially -- performative rather than deliberative \citep{Hannak.2014}. Furthermore, both approaches vary in terms of the effort required to fact\-/check {tweets
}. Writing a full\-/fledged community fact\-/check 
arguably requires more time and expertise. Hence, snoping may draw groups of fact\-/checkers that are less willing to invest the necessary efforts to write for a full\-/fledged community fact\-/check and/or select posts that are faster (or easier) to fact\-/check. In line with this notion, we also find that Snopers fact\-/check {tweets
} at a higher speed. In sum, our findings imply that different fact\-/checking approaches -- carried out on the same social media platform -- can result in vastly different social media posts getting fact\-/checked.

These findings have important implications for future research studying misinformation on social media. Previous research has predominantly identified misinformation based on the presence of replies linking to fact\-/checks from third\-/party fact\-/checking organizations -- \ie, based on snoping. For instance, many works have studied the diffusion patterns of misleading vs.\ not misleading (``snoped'') posts, finding that misinformation is more viral than the truth \citep[\eg,][]{Vosoughi.2018,Solovev.2022b,Friggeri.2014}. However, our analysis suggests that such an identification strategy may impede the generalizability of the findings. While we do not claim that the selection of users contributing to a dedicated community\-/based fact\-/checking system is more representative for the population of misinformation on social media as a whole, our results still imply that Notes contributors and Snopers focus on different targets when fact\-/checking social media content. Due to differences in user bases and content dynamics, earlier findings obtained for snoped posts might not apply to posts that have been fact\-/checked on community\-/based fact\-/checking systems such as Community Notes. Future research should be aware that sample selection plays a key role when studying misinformation and attempt to compile datasets that do not rely on a single fact\-/checking approach. In particular, compiling a representative sample of \emph{all} misinformation circulating on social media presents an important -- yet difficult -- challenge for future research.

From a practical perspective, our work has important implications for social media platforms, which can utilize our results to optimize community\-/based fact\-/checking systems and strategies to combat misinformation. The observed differences in the selection of fact\-/checking targets suggest that both approaches might complement each other well. Actively encouraging fact\-/checking of social media content via both snoping and dedicated community\-/based fact\-/checking systems (such as Community Notes) could lead to improved coverage and may help to combat misinformation on social media more effectively. Alternatively, platforms could integrate snopes on their platforms (\eg, by highlighting fact\-/checks in reply threads) or even actively encourage users that have snoped a social media post to write a full\-/fledged community fact\-/check. Platforms could further combine both approaches with machine learning, in order to enhance early warning systems for misinformation. In sum, by considering both snopes and Community Notes, future work might develop more effective strategies for reducing the proliferation of misinformation.

\textbf{Limitations and Future Research:} Our work has several limitations, which provide promising opportunities for future research. First, due to the observational nature of our work, we report associations and refrain from making causal claims. Second, more research is necessary to better understand {which} groups of users engage in community\-/based fact\-/checking and differences in their expertise. {Third, Twitter may have removed some particularly egregious misinformation through content moderation efforts. However, related work suggests that the number of deleted tweets is relatively small and unlikely to change the main findings in observational misinformation studies \cite{Solovev.2022b}.} Fourth, our inferences are limited to community\-/based fact\-/checking on the social media platform Twitter and the pilot phase of the Community Notes feature. Community\-/based fact\-/checking on Twitter may evolve to a different steady\-/state due to a growing/more experienced user base and changes in functionality. Future work may analyze whether the observed patterns are generalizable to posts from other fact\-/checking systems and social media platforms. Lastly, more research is necessary to better understand the role of manipulation attempts, (political) biases, performative vs.\ deliberative motivations, and the conditions under which the wisdom of crowds can be unlocked for fact\-/checking.

\section{Conclusion}

The spread of misinformation on social media is a pressing societal problem that researchers and practitioners continue to grapple with. As a countermeasure, recent research proposed to build on crowd wisdom to fact\-/check social media content. In this study, we empirically analyzed how community fact\-/checkers select their targets on social media. For this purpose, we compared the characteristics of social media posts that have been community fact\-/checked on Twitter's Community Notes platform with social media posts that have been snoped. Our analysis implies that Notes contributors and Snopers focus on different targets when fact\-/checking social media content and that both approaches might well complement each other. These findings have important implications for social media providers, which can use our results to optimize community\-/based fact\-/checking systems and strategies to combat misinformation on their platforms.

\section{Ethics Statement}
This research did not involve interventions with human subjects, and, thus, no approval from the Institutional Review Board was required by the authors' institutions. All analyses are based on publicly available data. To respect privacy, we explicitly do not publish usernames in our paper and only report aggregate results. We declare no competing interests. 

\section{Acknowledgments}
This study was supported by a research grant from the German Research Foundation (DFG grant 492310022).


\bibliography{literature}


\end{document}